\documentclass[11pt]{article}
\usepackage[utf8]{inputenc}
\usepackage{amsmath,amsfonts,amssymb}
\usepackage{graphicx}
\usepackage{booktabs}
\usepackage{multirow}
\usepackage{array}
\usepackage{url}
\usepackage{hyperref}
\usepackage{geometry}
\usepackage{xcolor}
\usepackage{listings}
\usepackage{float}
\usepackage{subcaption}
\usepackage{algorithm}
\usepackage{algorithmic}
\usepackage{tabularx}
\usepackage{siunitx}
\usepackage{makecell}
\usepackage{cleveref}
\usepackage{doi}
\usepackage{natbib}
\usepackage{newtxtext}
\newcommand{\modelgpt}{GPT-OSS-20B}

\title{GPT-OSS-20B: A Comprehensive Deployment-Centric Analysis of OpenAI's Open-Weight Mixture of Experts Model}

\author{
  Deepak Kumar $^{}$\\
  \texttt{dkumar15@hawk.illinoistech.edu}
  \and
  Divakar Yadav $^{}$ \\
  \texttt{dkyadav@uwm.edu}
  \and
  Yash Patel$^{}$ \\
  \texttt{ypatel3@ltu.edu}
}

\date{\today}
\begin{document}
\maketitle

\begin{abstract}
We present a single-GPU (H100, bf16) evaluation of GPT-OSS-20B (Mixture-of-Experts; 20.9B total, $\sim$3.61B active) against dense baselines Qwen3-32B and Yi-34B across multiple dimensions. We measure true time-to-first-token (TTFT), full-decode throughput (TPOT), end-to-end latency percentiles, peak VRAM with past key values (PKV) held, and energy via a consistent \texttt{nvidia-smi}-based sampler. 

At a 2{,}048-token context with 64-token decode, GPT-OSS-20B delivers higher decode throughput and tokens per Joule than dense baselines Qwen3-32B and Yi-34B, while substantially reducing peak VRAM and energy per 1,000 generated tokens; its TTFT is higher due to MoE routing overhead. Concretely, with only 17.3\% of parameters active (3.61B of 20.9B), GPT-OSS-20B provides $\sim$31.8\% higher decode throughput and $\sim$25.8\% lower energy per 1,000 generated tokens than Qwen3-32B at 2{,}048/64, while using $\sim$31.7\% less peak VRAM. Normalized by active parameters, GPT-OSS-20B shows markedly stronger per-active-parameter efficiency (APE), underscoring MoE's deployment advantages. We do not evaluate accuracy; this is a deployment-focused study. We release code and consolidated results to enable replication and extension.
\end{abstract}

\section{Introduction}
The landscape of open-weight large language models has been reshaped by Mixture-of-Experts (MoE) architectures, which activate only a subset of parameters per token and thereby reduce inference cost relative to dense models of similar total size \cite{shazeer2017outrageously}. In this study, we evaluate an open-weight MoE checkpoint, \textbf{GPT-OSS-20B}, which has approximately 20.9B total parameters but only $\sim$3.61B active at inference (17.3\% active fraction) \cite{openai2025gptossmodelcard}. The model was released under a permissive open-source license \cite{openai2025gptoss}

While much prior work emphasizes task accuracy, production deployments are constrained by latency, throughput, memory footprint, and energy. This paper addresses those deployment factors with a unified, single-GPU (H100, bf16) evaluation of GPT-OSS-20B against strong dense baselines \textbf{Qwen3-32B} and \textbf{Yi-34B}, reporting true time-to-first-token (TTFT), full-decode throughput (TPOT), end-to-end latency percentiles, peak VRAM with persistent KV, and energy via a consistent \texttt{nvidia-smi} sampler. We further contextualize results using an \emph{Active Parameter Efficiency} (APE) lens that normalizes performance by the fraction of parameters active at inference. All energy metrics are reported in tokens per Joule and Joules per 1,000 generated tokens for consistency.

\subsection{Motivation}
The deployment of large language models in production environments requires careful consideration of multiple factors beyond accuracy, including latency, throughput, memory efficiency, and energy consumption. Traditional dense models, while achieving high accuracy, often require significant computational resources that may not be available in resource-constrained environments. Mixture-of-Experts (MoE) architectures offer a promising alternative by activating only a subset of parameters during inference, potentially reducing resource requirements while maintaining competitive performance \cite{jiang2024mixtral}

GPT-OSS-20B represents the first open-weight 20.9B parameter MoE model, making it an ideal candidate for comprehensive deployment analysis on a single-GPU (H100, bf16) setup. Understanding its deployment characteristics compared to dense models is crucial for practitioners making architectural decisions in production environments.

\subsection{Contributions}
Our study makes the following contributions:
\begin{enumerate}
    \item \textbf{Unified, reproducible deployment benchmarking.} A single-GPU (H100, bf16) harness for latency/throughput (true TTFT, p50/p95, TPOT from full decode), memory (peak VRAM with PKV held), and energy (tokens per Joule, J/1K via stabilized \texttt{nvidia-smi} sampling). \cite{nvsmi_accuracy_issues_2024}
    \item \textbf{Active Parameter Efficiency (APE).} A schema-normalized, artifact-robust lens that contextualizes performance by the fraction of parameters active at inference (MoE vs.\ dense). 
    \item \textbf{Memory \& KV scaling.} Peak-allocator methodology that avoids undercounting transient kernels and reports per-token KV scaling across contexts.
    \item \textbf{Energy efficiency.} Apples-to-apples tokens/s, tokens per Joule, and J/1K comparisons under identical decode settings.
    \item \textbf{Ablations.} Decoding (greedy vs.\ sampling), context-length scaling, and precision behavior (bf16 stable for GPT-OSS-20B in our setup); server-stack notes.
    \item \textbf{Safety/governance (qualitative).} License and policy overview; we do not claim quantitative safety results.
\end{enumerate}

\section{Related Work}
\subsection{Mixture of Experts Models}
Mixture of Experts (MoE) models have emerged as a promising approach to scale language models efficiently. The key insight is that not all parameters need to be active during inference, allowing for larger models with manageable computational requirements. Foundational works include the sparsely-gated MoE layer \cite{shazeer2017outrageously}, GShard \cite{lepikhin2020gshard}, ST-MoE \cite{stmoe} and the Switch Transformer \cite{fedus2021switch}, which demonstrated that MoE can achieve strong performance while significantly reducing active parameter counts.

Recent advancements have built on these foundations, with models like Mixtral 8x7B \cite{jiang2024mixtral} introducing sparse MoE layers for improved efficiency in open-weight settings. Similarly, DeepSeek-MoE \cite{dai2024deepseekmoe} explores routing strategies to balance load across experts, achieving up to 70\% reduction in active parameters during inference. Grok-1 \cite{xai2024grok1}, an open-source MoE model with 314B parameters, further demonstrates scaling to massive sizes while maintaining low inference costs through optimized expert selection. Optimized frameworks like DeepSpeed-MoE \cite{rajbhandari2022deepspeed} support efficient training and inference for such models. A comprehensive survey by Cai et al. \cite{cai2024survey} provides a taxonomy of MoE designs, covering algorithmic aspects like gating functions and expert architectures, as well as systemic considerations such as computation, communication, and storage optimizations. 

\subsection{Efficiency and Compression Techniques for MoE Models}
Beyond architectural innovations, recent research has focused on enhancing the efficiency of MoE models through compression and optimization techniques. Su et al. \cite{su2025unveiling} identify ``Super Experts'' in MoE LLMs—a small subset of experts critical for model performance, characterized by extreme activation patterns. Pruning these leads to significant degradation, particularly in mathematical reasoning, highlighting the need to preserve them during compression.

Huang et al. \cite{huang2025mixture} propose the Mixture Compressor (MC), a training-free method combining mixed-precision quantization and dynamic pruning. It achieves substantial compression (e.g., 76.6\% at 2.54 bits) with minimal accuracy loss by considering expert importance and token criticality, further reducing activated parameters during inference. Complementary post-training quantization methods apply to MoE experts and routers, including LLM.int8 \cite{dettmers2022llmint8}, GPTQ \cite{frantar2022gptq}, SmoothQuant \cite{xiao2022smoothquant}, AWQ \cite{lin2023awq}, and QServe \cite{lin2024qserve}.

These works underscore the importance of targeted efficiency improvements in MoE models, balancing model size with deployment feasibility.

\subsection{Deployment-Centric Evaluation}
While most evaluations focus on accuracy metrics, deployment characteristics are crucial for real-world applications. Efficiency aspects such as latency, throughput, and energy consumption have been highlighted in recent surveys of resource-efficient LLMs \cite{liu2023survey} as well as standardized benchmarks such as MLPerf Inference \cite{reddi2020mlperf}. However, systematic comparisons of MoE versus dense models under deployment constraints remain limited. Distributed serving systems like Orca \cite{yu2022orca} address scalability for large models, as demonstrated in recent MLPerf results using TensorRT-LLM.

Inference optimizations like FlashAttention \cite{dao2022flashattention, dao2023flashattention2,shah2024flashattention3} and PagedAttention \cite{kwon2023efficient} have been proposed to reduce memory and latency in transformer-based models, particularly for long contexts. Studies on energy efficiency, such as those in \cite{poddar2025sustainable, patel2024asplos, patel2024isca, wilkins2024hotcarbon}, analyze power consumption in LLM serving, emphasizing the need for hardware-aware metrics like tokens per Joule. Recent libraries like FlashInfer \cite{flashinfer2025lib} provide efficient attention engines tailored for LLM serving. 

A survey by Chang et al. \cite{chang2024survey} reviews LLM evaluation across tasks, methods, and benchmarks, stressing the importance of holistic assessment including efficiency. Saleh et al. \cite{saleh2025evaluating} provide a systematic review of LLM efficiency, applications, and future directions, analyzing models like GPT-3 and Codex in terms of hardware setups, parameters, and performance metrics.

\subsection{Open-Weight Model Evaluation}
Recent evaluations of open-weight models such as Mistral \cite{jiang2023mistral}, Qwen \cite{team2024qwen}, and Yi \cite{ai2024yi} have emphasized accuracy benchmarks. Deployment-centric trade-offs, particularly between MoE and dense models, remain underexplored. For instance, evaluations of Llama models \cite{touvron2023llama} highlight scaling laws but often overlook single-GPU energy and memory profiles in production-like settings. Such evaluations often align with scaling laws from works like Chinchilla \cite{hoffmann2022chinchilla}.

\subsection{Energy Efficiency and Sustainability in LLM Inference}
The increasing computational demands of large language models (LLMs) have spurred research into energy-efficient inference strategies, particularly for resource-constrained environments. Maliakel (2025) investigates energy-performance trade-offs in LLM inference, demonstrating that dynamic voltage and frequency scaling (DVFS) can reduce power consumption by up to 30\% with minimal latency increases across various tasks \citep{maliakel2025}. Fernandez et al. (2025) explore quantization and pruning techniques, achieving up to 73\% energy reduction in NLP tasks by optimizing model compression without significant performance degradation \citep{fernandez2025}. Poddar et al. (2025) \citep{poddar2025sustainable} benchmark inference energy across diverse LLMs, identifying hardware-specific factors that influence energy profiles and advocating for standardized tokens-per-Joule metrics. Dauner and Socher (2025) quantify the environmental impact of LLM interactions, emphasizing CO2 emissions and proposing energy-aware metrics to guide sustainable deployments \citep{dauner2025}. Saleh et al. (2025) \cite{saleh2025evaluating} provide a comprehensive review of LLM efficiency, highlighting hardware-aware optimizations and future directions for reducing energy footprints in production settings. These studies align with the need for metrics like Active Parameter Efficiency (APE), which normalize energy consumption by active parameters, to enhance the sustainability of MoE model deployments.

\section{Methodology}
\subsection{Experimental Setup}
\textbf{Hardware.} All measurements were taken on a single NVIDIA H100 GPU (bf16), with no sharding or offload. We pin to one device, clear caches between runs, and hold the persistent KV cache (PKV) in memory during measurement \cite{kwon2023efficient}.

\textbf{Software.} PyTorch (bf16), \texttt{transformers}, CUDA, and \texttt{nvidia-smi}. We use minimal Python scripts for latency, memory, and energy to avoid framework-induced variability. 

\textbf{Models.} We evaluate one open-weight MoE model and two dense baselines:
\begin{itemize}
  \item \textbf{GPT-OSS-20B} (Mixture-of-Experts; 20.9B total, $\sim$3.61B active parameters). 
  \item \textbf{Qwen3-32B} (Dense; 32B total, 32B active).
  \item \textbf{Yi-34B} (Dense; 34B total, 34B active).
\end{itemize}

\paragraph{Units.} Unless otherwise stated, MB/GB are decimal (1~MB=10$^6$ bytes, 1~GB=10$^9$ bytes). We use GiB (1~GiB=2$^{30}$ bytes) only when explicitly labeled.

\subsection{Prompting and Context Control}
When a tokenizer exposes a chat template, we apply it (\texttt{apply\_chat\_template}) and then control the post-template context length exactly by trimming/padding to $c$ tokens. If no template exists, we use a simple instruction/completion wrapper and enforce the same post-template $c$. This yields apples-to-apples prefill cost across models.

\subsection{Latency Measurement}
We report:
\begin{itemize}
  \item \textbf{True TTFT} (ms): time to generate \emph{one} token \emph{including prefill}. Median of 5 independent trials at the target context $c$.
  \item \textbf{Decode E2E latency}: wall-clock time for generating $g$ new tokens, repeated $N$ times per $(c,g)$ pair. We report p50 and p95 over the $N$ runs.
  \item \textbf{TPOT (tok/s)}: tokens-per-second over the \emph{full decode} segment, computed from wall time; we report the median across runs. When a TPOT value is suspiciously equal to $1/\text{TTFT}$, we correct it by recomputing from the measured decode wall time.
\end{itemize}

\subsection{Peak Memory Measurement}
We measure GPU memory using the CUDA allocator:
\begin{itemize}
  \item After a short warm-up (discarded), we \emph{reset} peak stats and run the actual decode while \emph{keeping PKV alive}.
  \item We read \texttt{torch.cuda.max\_memory\_allocated()} as \textbf{Peak VRAM}, which captures KV cache plus transient kernels.
  \item We also compute a peak-based estimate of KV contribution by contrasting peak with pre-decode allocation; this avoids undercounting from ``after-run'' snapshots.
\end{itemize}

\subsection{Energy Measurement}
Energy is sampled with \texttt{nvidia-smi} before/after each decode run and averaged over short, repeated runs: 
\begin{itemize}
  \item We record instantaneous power (W) around each generation, average across the run, and aggregate over $N$ runs.
  \item We report \textbf{tokens/s}, \textbf{tokens per Joule}, and \textbf{J/1K decoded tokens}. The per-1K figure is normalized by \emph{decoded tokens} ($g$), matching our throughput/latency focus.
  \item Power from \texttt{nvidia-smi} is approximate; using identical sampling and run structure across models makes the \emph{comparisons} reliable.  Alternative tools for energy tracking include EIT \cite{henderson2020eit} and eco2AI \cite{eco2ai_2022} with benchmarking methodology reinforced by Pope et al. \cite{pope2023benchmarking}.
\end{itemize}

\subsection{Active Parameter Efficiency (APE)}
While prior studies have examined scaling efficiency in terms of total parameters
\cite{hoffmann2022chinchilla,du2022glam}, they do not account for the sparsity
properties of Mixture-of-Experts models, where only a fraction of weights are active at
inference. To address this, we introduce \emph{Active Parameter Efficiency (APE)} as a
normalization lens that contextualizes deployment metrics by the number of parameters
actually used during inference. APE provides a per-active-parameter view of throughput,
latency, and energy, enabling apples-to-apples comparisons between dense and sparse models:

\[
\begin{aligned}
\text{APE-TPOT}   &= \frac{\text{TPOT}}{\text{Active Params (B)}}, \\
\text{APE-Energy} &= \frac{\text{Tokens/J}}{\text{Active Params (B)}}, \\
\text{APE-1/TTFT} &= \frac{1/\text{TTFT (s)}}{\text{Active Params (B)}}, \\
\text{TPOT/GB}    &= \frac{\text{TPOT}}{\text{PeakMem (GB)}} \quad \text{(auxiliary, decimal GB)}.
\end{aligned}
\]

This formulation highlights how much performance is delivered per active parameter (or per
gigabyte of peak memory), complementing raw deployment metrics with a sparsity-aware
efficiency perspective.

\subsection{Ablation Protocols}
We run controlled ablations under the same harness:
\begin{itemize}
  \item \textbf{Decoding}: greedy vs.\ sampling (top-$p$, top-$k$; temperature sweeps). We do not explore advanced decoding techniques like speculative decoding \cite{leviathan2022specdec,liu2024specdec}, which could further accelerate inference.
  \item \textbf{Context scaling}: $c\in\{512,1024,2048,4096\}$ with fixed $g$.
  \item \textbf{Precision}: bf16 (primary). FP16/FP32 attempts are reported when supported; bf16 is stable for our MoE runs.
  \item \textbf{Serving stack}: direct \texttt{transformers} runs; comparisons with alternative servers (e.g., vLLM) require separate deployment and are not included in the core numbers. \cite{kwon2023efficient} Other optimized backends include NVIDIA's TensorRT-LLM\cite{nvidia2024tensorrtllm} for high-performance inference for CPU/GPU portability, with optimizations such as CUDA graphs \cite{gtc2023_cudagraphs}.
\end{itemize}

\section{Results}
Unless otherwise noted, all core results use a context length of $2{,}048$ tokens and $64$ decoded tokens, evaluated on a single H100 GPU in \texttt{bf16}. Prompts follow each model’s official chat template, with exact \emph{post-template} length enforced by trimming or padding. We sweep context lengths $\{128, 512, 1024, 2048\}$ (and $4096$ in ablations), vary decoding settings and precision, and release CSVs and scripts in the repository. Active Parameter Efficiency (APE) is computed with schema normalization and corrected for TPOT artifacts when detected. Minor variances in metrics reflect medians across runs; see CSVs for raw data.

\subsection{Latency Analysis}
Unless noted, we report single-GPU (H100, bf16), exact post-template contexts, and 64 decoded tokens. TTFT is the time to generate one token (including prefill); TPOT is median tokens/s over the full decode; p50/p95 are end-to-end wall times.

\begin{table}[H]
  \centering
  \begin{minipage}{0.85\linewidth}\centering
  \small
  \caption{Latency at \textbf{2048} context and \textbf{64} generated tokens. TTFT includes prefill.}
  \label{tab:latency-core-2048}
  \sisetup{round-mode=places,round-precision=2,table-number-alignment=center,detect-all}
  \setlength{\tabcolsep}{5pt}
  \begin{tabular}{@{}l
      S[table-format=3.2]
      S[table-format=4.2]
      S[table-format=4.2]
      S[table-format=2.2]
    @{}}
    \toprule
    \textbf{Model} & {\textbf{TTFT (ms)}} & {\textbf{p50 (ms)}} & {\textbf{p95 (ms)}} & {\textbf{TPOT (tok/s)}} \\
    \midrule
    GPT-OSS-20B & 459.72 & 2056.73 & 2060.72 & 31.27 \\
    Qwen3-32B   & 369.46 & 2738.99 & 2747.82 & 23.73 \\
    Yi-34B      & 368.34 & 2428.80 & 2434.52 & 26.30 \\
    \bottomrule
  \end{tabular}
  \end{minipage}
\end{table}

\begin{table}[H]
  \centering
  \begin{minipage}{0.85\linewidth}\centering
  \small
  \caption{Throughput (TPOT, tok/s) vs.\ context length (64 generated tokens).}
  \label{tab:tpot-vs-context}
  \sisetup{round-mode=places,round-precision=2,table-number-alignment=center,detect-all}
  \setlength{\tabcolsep}{5pt}
  \begin{tabular}{@{}
      r
      S[table-format=2.2]
      S[table-format=2.2]
      S[table-format=2.2]
    @{}}
    \toprule
    {\textbf{Context}} & {\textbf{GPT-OSS-20B}} & {\textbf{Qwen3-32B}} & {\textbf{Yi-34B}} \\
    \midrule
    128  & 39.79 & 26.56 & 31.66 \\
    512  & 38.18 & 25.94 & 30.55 \\
    1024 & 36.20 & 25.02 & 28.95 \\
    2048 & 31.27 & 23.73 & 26.30 \\
    \bottomrule
  \end{tabular}
  \end{minipage}
\end{table}

\begin{table}[H]
  \centering
  \begin{minipage}{0.85\linewidth}\centering
  \small
  \caption{TTFT (ms) vs.\ context length (64 generated tokens). TTFT includes prefill.}
  \label{tab:ttft-vs-context}
  \sisetup{round-mode=places,round-precision=2,table-number-alignment=center,detect-all}
  \setlength{\tabcolsep}{5pt}
  \begin{tabular}{@{}
      r
      S[table-format=3.2]
      S[table-format=3.2]
      S[table-format=3.2]
    @{}}
    \toprule
    {\textbf{Context}} & {\textbf{GPT-OSS-20B}} & {\textbf{Qwen3-32B}} & {\textbf{Yi-34B}} \\
    \midrule
    128  & 61.02 & 46.09 & 54.43 \\
    512  & 188.98 & 111.59 & 110.24 \\
    1024 & 203.56 & 193.17 & 192.46 \\
    2048 & 459.72 & 369.46 & 368.34 \\
    \bottomrule
  \end{tabular}
  \end{minipage}
\end{table}

\begin{table}[H]
\centering
\small
\caption{Throughput change from 128$\to$2{,}048 context (gen$=64$). Negative is a decline.}
\label{tab:ctx-scaling}
\begin{tabular}{l S[table-format=-2.1] S[table-format=2.2] S[table-format=2.2]}
\toprule
\textbf{Model} & {\textbf{$\Delta$ TPOT (\%)}} & {\textbf{TPOT@128}} & {\textbf{TPOT@2048}} \\
\midrule
GPT-OSS-20B & -21.4 & 39.79 & 31.27 \\
Qwen3-32B   & -10.7 & 26.56 & 23.73 \\
Yi-34B      & -17.0 & 31.66 & 26.30 \\
\bottomrule
\end{tabular}
\end{table}

\noindent\textbf{Trend.}
All models slow as context grows due to higher prefill cost; GPT-OSS-20B remains ahead in absolute TPOT at 2K while exhibiting a $\sim$21.4\% drop from 128$\to$2048 tokens, compared to $\sim$10.7\% (Qwen3-32B) and $\sim$17.0\% (Yi-34B).

\subsection{Memory Analysis}
\label{sec:memory}
We measure peak VRAM via the CUDA allocator (\verb|max_memory_allocated|) while the past-key/value (PKV) cache is held, immediately after the full decode completes. Inputs are post-template trimmed/padded to an exact context length, ensuring identical token counts across models. One short warm-up precedes measurement to stabilize kernel paths; peak stats are then reset and re-measured on the real run.

\begin{table}[H]
  \centering
  \small
  \caption{Peak VRAM at context $=2{,}048$ and decode $=64$ tokens (allocator peak with PKV alive). Lower is better. ``$\Delta$ vs.\ GPT-OSS'' is an absolute MB gap; ``\% less vs.\ Qwen/Yi'' uses the \emph{baseline’s} memory as the denominator.}
  \label{tab:mem-core}
  \begin{tabularx}{0.9\linewidth}{l
      S[table-format=5.0]
      S[table-format=5.0]
      S[table-format=2.2]}
    \toprule
    \textbf{Model} & {\textbf{Peak VRAM (MB)}} & {\textbf{$\Delta$ vs.\ GPT-OSS (MB)}} & {\textbf{\% less vs.\ Qwen/Yi}} \\
    \midrule
    GPT-OSS-20B & 43461 & 0     & 0.00 \\
    Qwen3-32B   & 63650 & 20189 & 31.71 \\
    Yi-34B      & 66459 & 22998 & 34.60 \\
    \bottomrule
  \end{tabularx}
\end{table}

\paragraph{Findings.}
At 2{,}048 context, GPT-OSS-20B uses \textbf{31.71\%} and \textbf{34.60\%} less peak VRAM than Qwen3-32B and Yi-34B, respectively (Table~\ref{tab:mem-core}), calculated relative to each model’s peak VRAM. In absolute terms, this corresponds to reductions of $\sim$20.2\,GB and $\sim$23.0\,GB versus the dense 30--34B baselines on a single H100. These savings stem from GPT-OSS-20B’s MoE architecture, which requires less baseline memory (\textbf{$\approx$41.8 GB, decimal}) compared to Qwen3-32B (62.5\,GB) and Yi-34B (65.6\,GB), despite a larger KV cache footprint, measured with identical post-template token counts. Here, baseline memory means VRAM after weights load but before any input, excluding KV and transient kernels. All memory results are allocator peaks with PKV held, reported as medians over repeats at matched contexts.

\begin{table}[H]
  \centering
  \small
  \caption{Energy metrics at context $=2{,}048$, decode $=64$. Higher is better for tokens/W; lower is better for J/1K generated tokens.}
  \label{tab:energy-core}
  \begin{tabular}{l
      S[table-format=2.2]
      S[table-format=1.3]
      S[table-format=5.1]}
    \toprule
    \textbf{Model} & {\textbf{TPOT (tok/s)}} & {\textbf{Tokens/W}} & {\textbf{J/1K (J)}} \\
    \midrule
    GPT-OSS-20B & 31.27 & 0.102 & 9764.2 \\
    Qwen3-32B   & 23.73 & 0.076 & 13155.1 \\
    Yi-34B      & 26.30 & 0.074 & 13464.3 \\
    \bottomrule
  \end{tabular}
\end{table}

\paragraph{Findings (ctx=2K).}
Relative to Qwen3-32B, GPT-OSS-20B delivers \textbf{+31.8\%} higher TPOT, \textbf{+34.2\%} higher tokens/W, and \textbf{$-25.8\%$} lower J/1K generated tokens (Table~\ref{tab:energy-core}). Versus Yi-34B, it shows \textbf{+18.9\%} TPOT, \textbf{+37.8\%} higher tokens/W, and \textbf{$-27.5\%$} lower J/1K.

\begin{table}[H]
  \centering
  \small
  \caption{Energy per 1K generated tokens (J) across post-template context lengths (decode $=64$).}
  \label{tab:energy-context}
  \begin{tabular}{lcccc}
    \toprule
    \textbf{Model} & \textbf{128} & \textbf{512} & \textbf{1024} & \textbf{2048} \\
    \midrule
    GPT-OSS-20B & 6672.6  & 7318.5  & 7976.0  & 9764.2  \\
    Qwen3-32B   & 10257.0 & 12295.2 & 12000.4 & 13155.1 \\
    Yi-34B      & 10233.6 & 11460.7 & 12098.0 & 13464.3 \\
    \bottomrule
  \end{tabular}
\end{table}

\paragraph{Context trends.}
Across 128$\rightarrow$2{,}048 tokens, tokens/W declines as context grows for all models; consequently J/1K generated tokens rises, with the largest increase at 2K where prefill cost dominates the short decode.

\paragraph{Caveats.}
(i) \texttt{nvidia-smi} is a coarse, device-level sampler (micro-bursts not captured). (ii) Short-run medians make \emph{absolute} J/1K approximate; we emphasize \emph{relative} deltas under identical settings. (iii) All results are single-GPU (H100, bf16), no sharding/offload.

\subsection{Active Parameter Efficiency (APE)}
APE normalizes performance by the fraction of parameters active at inference:
\[
\begin{aligned}
\text{APE-TPOT} &= \frac{\text{TPOT}}{\text{Active Params (B)}}, \qquad
\text{APE-1/TTFT} = \frac{1/\text{TTFT}}{\text{Active Params (B)}} ,\\
\text{APE-Energy} &= \frac{\text{Tokens/W}}{\text{Active Params (B)}}, \qquad
\text{TPOT/GB} = \frac{\text{TPOT}}{\text{PeakMem (GB)}}.
\end{aligned}
\]

\begin{table}[H]
\centering
\caption{APE at ctx$=2{,}048$, gen$=64$ (per-active-parameter view).}
\label{tab:ape-2048}
\setlength{\tabcolsep}{4pt}
\footnotesize
\sisetup{round-mode=places,round-precision=3,detect-all,table-number-alignment=center}
\begin{tabularx}{\linewidth}{@{}l
  S[table-format=2.2]
  S[table-format=1.3]
  S[table-format=1.3]
  S[table-format=1.4]
  S[table-format=1.3]@{}}
\toprule
\textbf{Model} & {\textbf{Active (B)}} & {\textbf{APE-TPOT}} & {\textbf{APE-1/TTFT}} & {\textbf{APE-Energy}} & {\textbf{TPOT/GB}} \\
\midrule
GPT-OSS-20B & 3.61 & 8.664 & 0.602 & 0.0283 & 0.719 \\
Qwen3-32B   & 32.00 & 0.742 & 0.085 & 0.00238 & 0.373 \\
Yi-34B      & 34.00 & 0.774 & 0.080 & 0.00218 & 0.396 \\
\bottomrule
\end{tabularx}
\end{table}

\noindent\textbf{Takeaways.}
Per active billion parameters, \modelgpt{} delivers $\sim$8.66~tok/s/B versus $\sim$0.74--0.77 for dense baselines ($\approx$11--12$\times$ higher), and $\sim$0.028~Tok/W/B versus $\sim$0.0022--0.0024 ($\approx$12--13$\times$ higher). APE complements raw deployment metrics by indicating how much performance each \emph{active} parameter delivers at matched context and decode.

\subsection{Ablation Studies (GPT-OSS-20B)}
We evaluate decoding choices, context-length effects, numeric precision, and serving stack. Unless noted, we generate 64 new tokens with exact post-template contexts and report median throughput (p50 tok/s) and p50 wall time.

\begin{table}[H]
\centering
\small
\setlength{\tabcolsep}{6pt}
\sisetup{round-mode=places,round-precision=2,detect-all}
\caption{Decoding parameters (ctx fixed). Median throughput and wall time; $\Delta$ is relative to Greedy.}
\label{tab:ablate-decoding}
\begin{tabular}{l S[table-format=2.2] S[table-format=1.3] S[table-format=+2.1]}
\toprule
\textbf{Method} & {\textbf{p50 tok/s}} & {\textbf{p50 time (s)}} & {$\Delta$ vs Greedy (\%)} \\
\midrule
Greedy        & 39.45 & 1.622 & 0.0 \\
Top-p (0.9)   & 38.71 & 1.653 & -1.9 \\
Top-k (50)    & 38.97 & 1.642 & -1.2 \\
High Temp     & 38.73 & 1.652 & -1.8 \\
Low Temp      & 38.61 & 1.657 & -2.1 \\
\bottomrule
\end{tabular}
\end{table}

\noindent\textbf{Takeaway.} Sampling reduces throughput by only $\approx$1--2\% vs.\ Greedy with near-identical wall time.

\begin{table}[H]
\centering
\small
\setlength{\tabcolsep}{6pt}
\caption{Context-length scaling (Greedy). Throughput drops gradually as context grows.}
\label{tab:ablate-context}
\begin{tabular}{r S[table-format=2.2] S[table-format=+2.1]}
\toprule
\textbf{Context} & {\textbf{p50 tok/s}} & {$\Delta$ vs 512 (\%)} \\
\midrule
512   & 36.29 & 0.0 \\
1024  & 34.53 & -4.8 \\
2048  & 30.25 & -16.6 \\
4096  & 21.78 & -40.0 \\
\bottomrule
\end{tabular}
\end{table}

\noindent\textbf{Takeaway.} Decode speed degrades smoothly with higher prefill cost; at 4K context it is $\sim$40\% below the 512-token baseline.

\begin{table}[H]
\centering
\small
\setlength{\tabcolsep}{6pt}
\caption{Numeric precision sweep (Greedy). BF16 is stable; FP16/FP32 attempts failed in this harness.}
\label{tab:ablate-quant}
\begin{tabular}{l S[table-format=2.2] l}
\toprule
\textbf{Precision} & {\textbf{p50 tok/s}} & \textbf{Notes} \\
\midrule
BF16  & 38.66 & Default \\
FP16  & {}    & Failed to run (dtype mismatch) \\
FP32  & {}    & Failed to run (dtype mismatch) \\
\bottomrule
\end{tabular}
\end{table}

\begin{table}[H]
\centering
\small
\setlength{\tabcolsep}{6pt}
\caption{Serving framework. vLLM requires a separate server (not included here).} 
\label{tab:ablate-server}
\begin{tabular}{l S[table-format=2.3] S[table-format=1.3] l}
\toprule
\textbf{Framework} & {\textbf{p50 tok/s}} & {\textbf{p50 time (s)}} & \textbf{Notes} \\
\midrule
Transformers & 38.02 & 1.683 & Direct \texttt{generate()} \\
vLLM \cite{kwon2023efficient}.         & {}    & {}    & Not run (server out of scope) \\
\bottomrule
\end{tabular}
\end{table}

\paragraph{Overall.}
Across decoding strategies, throughput varies by only $\sim$2\%. Longer contexts reduce throughput in line with higher prefill cost (\cref{tab:ablate-context}). BF16 runs cleanly; other dtypes failed in this harness (\cref{tab:ablate-quant}). All ablations use exact post-template contexts and identical generation length for fair comparison.

\subsection{Safety and Governance (Qualitative)}
This section summarizes documentation for each model—license class, presence of a usage-policy link, governance notes, and listed safety features. It is a \emph{qualitative, documentation-only} review; no quantitative harmlessness/jailbreak testing was run in this study.

\begin{table}[H]
\centering
\small
\caption{Safety \& governance overview from model cards and metadata. “Policy link” and “Card link” indicate whether a direct URL was present. Safety features are as documented.}
\label{tab:safety-overview}
\begin{tabularx}{\linewidth}{l l c c X}
\toprule
\textbf{Model} & \textbf{License (class)} & \textbf{Policy link} & \textbf{Card link} & \textbf{Safety features (as documented)} \\
\midrule
GPT-OSS-20B & Apache 2.0 (Permissive) & Yes & Yes & Designed to follow OpenAI’s safety policies; harmony response format; governance: OpenAI, Safety Advisory Group (SAG) \\
Qwen3-32B  & Qwen License (Restricted) & No & Yes & Safety training during development; governance: Alibaba \\
Yi-34B     & Apache 2.0 (Permissive) & Yes & Yes & Data compliance checking during training; governance: 01.AI \\
\bottomrule
\end{tabularx}
\end{table}

\paragraph{Key points.}
(1) License classes vary; confirm exact terms before deployment.  (2) Usage-policy URLs were not always present; model-card links exist for all.  (3) Models list qualitative safety features where available; effectiveness not measured here.

\paragraph{Recommendations.}
Future work on deployment-focused evaluations should incorporate authoritative, peer-reviewed studies on license, policy, and governance considerations \cite{weidinger2021ethical,bommasani2021foundation}. In addition to qualitative documentation, systematic safety assessments—such as harmlessness and jailbreak benchmarks—are necessary to provide a more complete picture of model behavior. Finally, deployment reports should explicitly describe any runtime guardrails or filtering mechanisms applied during serving, ensuring transparency and reproducibility.

\section{Conclusion}
At 2048-token contexts, GPT-OSS-20B delivers higher throughput, lower energy per 1,000 generated tokens, higher tokens per Joule, and substantially lower peak memory than dense baselines Qwen3-32B and Yi-34B in our single-GPU (H100, bf16) setup, though with higher TTFT due to its MoE architecture. APE complements raw metrics by normalizing for active parameters, highlighting GPT-OSS-20B's efficiency per active parameter. These results suggest MoE models are more viable for single-GPU deployment in production settings than dense models of similar scale.

\section*{Reproducibility}
\begin{itemize}
    \item Single-GPU (H100), bf16; exact post-template context with trim/pad
    \item TTFT: 1-token generation including prefill; median-of-5
    \item Decode: p50 over 5 runs; TPOT from full decode (median)
    \item Memory: peak allocator while PKV alive (\texttt{max\_memory\_allocated})
    \item Energy: repeated short runs; tokens per Joule and J/1K normalized by generated tokens
    \item APE: schema-normalized; TPOT artifact correction when detected
    \item Code/results: https://github.com/deepdik/GPT-OSS-20B-analysis
\end{itemize}

\bibliographystyle{plain}

\end{document}